\documentclass[aps,prd,twocolumn,showpacs,nofootinbib,amsmath,amssymb]{revtex4}

\newcommand{\be}{\begin{eqnarray}}
\newcommand{\ee}{\end{eqnarray}}
\newcommand{\nn}{\nonumber\\}

\newcommand{\I}{\mathrm{i}}
\newcommand{\e}{\mathrm{e}}
\newcommand{\tr}{\mathrm{tr}}

\newcommand{\SSH}{\hskip -1mm\not\hskip -1mm}

\newcommand{\soint}{{\textstyle\oint}}
\newcommand{\sfrac}[2]{\textstyle\frac{#1}{#2}}
\newcommand{\sint}{{\textstyle\int}}

\usepackage{bbm}

\begin{document}
\title
{On invariants for particle propagation in non-Abelian fields}
\author{Dennis D.~Dietrich}
\affiliation{Centre for Particle Physics Phenomenology (CP$^3$-Origins), University of Southern Denmark, Odense, Denmark}
\date{August 29, 2009}
\begin{abstract}
Characterising the propagation of particles in an external non-Abelian field
only in terms of invariants constructed from its field tensor is not
always sufficient, especially, in many analytically tractable and phenomenologically interesting
cases.
\end{abstract}
\maketitle


The concept of external fields has many uses, from theoretical
tools 
to phenomenological motivations. In the
latter case, the motivations range from computational feasibility to the
fact that the vacuum without external field is not the correct expansion
point \cite{DDD:ConstantFields}.
The investigation of quantum electrodynamics in external fields leads, e.g.,
to the seminal and as yet untested prediction of particle creation
in (originally constant electric) external fields \cite{ConstantFieldHistory}.
This and other effects are about to be tested, e.g., with ultra strong light 
sources \cite{ELI}. The generalisation to quantum chromodynamics is of interest 
in the context of high energy collisions. A concept that
is inseparably linked to external fields is that of effective actions 
\cite{ConstantFieldHistory,Dunne:2004nc}. For a covariantly constant background [see Eq.~(\ref{covconst})],
the corresponding computations proceed in close analogy to the Abelian case \cite{nabeffact}.

Observables in gauge field theories are by definition gauge invariant.
In the presence of backgrounds this means that the results may only depend
on said background in a gauge invariant way. A way to make the gauge
invariance manifest is to identify gauge invariant combinations of the background
field tensor \cite{Invariants} and express the observables in terms of these.
Backgrounds allowing for analytically tractable calculations, due
to technical limitations, have typically only a small number of nonzero 
Lorentz and colour components. Therefore, they are subject to the 
Wu--Yang ambiguity \cite{WY}. It states that in non-Abelian field theories 
there exist field tensors that have realisations in terms of different gauge
field configurations that are {\it not} gauge equivalent.
To see that 
these different gauge fields do indeed lead to different physics
consider the constant non-Abelian field tensor,
\be
E_3^a=F_{03}^a=\partial_0A^a_3-\partial_3A_0^a+f^{abc}A^b_0A^c_3, 
\ee
and all other components equal to zero. $f^{abc}$ stands for the antisymmetric 
structure constant of the gauge group $\mathcal{G}$. This field tensor can be realised by the gauge field
\be
A_3^a=+E_3^ax^0 ,  
\label{config3}
\ee
and zero otherwise. The gauge transformation
$
U=\e^{-\I E_3x^3x^0} ,
$
where $E_3=E_3^aT^a$ and $T^a$ represent the generators of $\mathcal{G}$, 
turns it into
$
A_0^a=-E_3^ax^3 ,
$
while leaving the field tensor invariant. Now regard
\be
A_0^a=a_0^a~~~~~\mathrm{and}~~~~~A_3^a=a_3^a,
\label{config03}
\ee
where $a_0^a$ and $a_3^a$ are constant such that $f^{abc}a^b_0
a^c_3=E_3^a$. (All of the above field configurations satisfy Lorenz as 
well as Coulomb gauge.) 
The gauge transformation that removes $a_3^a$ reads
$
U=\e^{-\I a_3x^3}.
$
It turns $a_0$ into
\be
Ua_0U^\dagger
=
\e^{-\I a_3x^3}a_0\e^{+\I a_3x^3}
=
a_0\e^{+2\I a_3x^3}
\neq 
a_0 ,
\ee
where we assumed $\{a_0,a_3\}=0$ for simplicity.
This gauge transformation also does not leave the field 
tensor invariant: Assuming $\{a_3,E_3\}=0$,
\be
UE_3U^\dagger=E_3\e^{+2\I a_3x^3} .
\ee

Another way of seeing that this last configuration is not gauge equivalent to 
the first is computing gauge invariant Wilson loops. Take the rectangular path
$\mathcal{C}$
$(x^0,x^3):(0,0)\rightarrow(y^0,0)\rightarrow(y^0,y^3)\rightarrow(0,y^3)\rightarrow(0,0)$.
For configurations (\ref{config3}) and (\ref{config03}) this yields
\be
W
&=&
\tr\e^{\I\soint_\mathcal{C}dx\cdot A}
=
\tr~\e^{\I E_3y^0y^3} \mathrm{~~~and}
\label{wilson0}\\
W
&=&
\tr~\e^{-\I a_3y^3}\e^{-\I a_0y^0}\e^{\I a_3y^3}\e^{\I a_0y^0} ,
\label{wilson03}
\ee
respectively, which do not coincide.

In 4 dimensions, a necessary condition for the presence of this ambiguity
is $\det\mathbbm F=0$, where
$\mathbbm{F}^{ab}_{\mu\nu}=\sfrac{1}{2}\epsilon_{\mu\nu\kappa\lambda}F^{c\kappa\lambda}f^{abc}$
\cite{WY:Det4d}. 
(Accordingly, such a determinant also appears as part of the Jacobian when 
translating path integrals from the gauge field to a field tensor 
formulation \cite{B}.)
$\mathbbm{F}$ is in the adjoint 
representation. Therefore, each submatrix of a single Lorentz component has 
zero eigenvalues. The corresponding eigenvectors of different submatrices
must be misaligned to have $\det\mathbbm{F}\neq0$.

Further, configuration (\ref{config3}) is covariantly constant 
(a gauge invariant statement), 
\begin{equation}
D_\lambda F_{\mu\nu}=0~~\forall~~\lambda,\mu,\nu ,
\label{covconst}
\end{equation}
as there 
$
D_\lambda F_{\mu\nu}=\partial_\lambda F_{\mu\nu}=0~~\forall~~\lambda,\mu,\nu .
$
For configuration (\ref{config03}) we have
$
\partial_\lambda F_{\mu\nu}=0~~\forall~~\lambda,\mu,\nu ,
$
and thus,
$ 
D_\lambda^{cd} F_{\mu\nu}^d
=
f^{abc}A^a_\lambda F^b_{\mu\nu}
\neq
0
$
for $\lambda,\mu,\nu\in\{0;3\}$. 

Thus, here $D_\lambda F_{\mu\nu}$ are the gauge covariant quantities that allow us
to distinguish between the gauge-inequivalent settings. They cannot be
expressed in terms of $F^a_{\mu\nu}$ alone. They can serve to
construct gauge invariant quantities, which can also be contracted into
Lorentz scalars. In particular, the current in the Yang--Mills equation,
$D_\mu^{ab}F^{b\mu\nu}=J^{a\nu}$, can be used in $J^a_\mu J^{a\mu}$. After 
all, covariant conservation is a sufficient albeit not necessary condition 
for a vanishing current. Hence, for covariantly constant fields all 
the invariants involving $J^a_\mu$ are zero. A related invariant is
$(D_\kappa^{ab}F_{\mu\nu}^b)(D^{ac\kappa}F^{c\mu\nu})$ \cite{Brown:1979bv}. 

In fact, the covariant derivative and not the field 
tensor is the elementary building block for invariants, in the sense that it
carries more information than the latter. Odd powers of the covariant 
derivative cannot be contracted into Lorentz scalars. Order 2 does not have 
nontrivial contributions. Order 4 has $F_{\mu\nu}F^{\mu\nu}$ and 
$F_{\mu\nu}\tilde F^{\mu\nu}$. Order 6 contains the aforementioned 
$J^a_\mu J^{a\mu}$. 

A rescaling $a_0\rightarrow a_0c$, $a_3\rightarrow a_3/c$
leaves the field tensor invariant \cite{Brown:1979bv}. 
This rescaling cannot be generated by a unitary global gauge transformation
and hence, the parameter $c$ characterises a continuous class of 
gauge-inequivalent gauge field configuration belonging to the same field
tensor. (There are no additional classes of gauge-inequivalent
representations, for a constant field tensor; covariantly constant and static 
configurations exhaust all possibilities \cite{Brown:1979bv}.) 
Fixing as reference $J^a_\mu J^{a\mu}=0$ for $c=1$, we obtain
$J^a_\mu J^{a\mu}=
(c^{-2}-c^2)|E_3|^3$. The $c=1$ case can be told apart from $J^a_\mu=0$
by means of $J^a_\mu J^{b\mu}J^a_\nu J^{b\nu}=(c^{-4}+c^4)|E_3|^6$, where
for simplicity we assumed $J_0^aJ_3^a=0$.
After inclusion of a third gauge field component, 
such that all components are noncommuting, which leads 
to 3 nonzero components for the field tensor, 
this continuous scaling symmetry breaks down to a simultaneous overall sign 
change.
The Wilson loop (\ref{wilson03}) is also $c$ dependent. In comparison, 
Klein--Gordon and Dirac propagators have additional structure 
\cite{DDD:Propagators}. To illustrate more how much the situations with
equal field tensor, but different gauge-inequivalent gauge fields
differ, we study these propagators in the presence of the 2
different configurations (\ref{config3}) and (\ref{config03}).

A {\it more general nonstatic configuration} $E_3=E_3(x^0)=[Q,E(x^0)]$, where
$Q=Q^aT^a=$constant, can also be realised either as a derivative of
\be 
A_3=A_3(x^0)=[Q,\sint dx^0E(x^0)] ,
\label{config3t}
\ee
or as a commutator of a constant
\be
A_3=QC~~~~~\mathrm{with}~~~~~A_0=E(x^0)/C,
\label{config03t}
\ee 
where $C$ also accounts for the correct mass dimension of the vector
potential. Clearly, $\det\mathbbm{F}=0$.
These field configurations still satisfy Coulomb, but not always Lorenz
gauge because of $A_0=A_0(x^0)$, which, however, could be rotated away.
For a covariantly constant electric field, the first realisation leads only to
$J_3\neq 0$, while the second has also $J_0\neq0$, i.e., a net charge
density. For the second configuration, the field tensor can again not
distinguish between gauge fields rescaled by a constant $c$ as described
above, which here is equivalent to dividing $C$ by $c$.
When it comes to gauge transformations, $A_3=QC$ can be removed by 
$U=\e^{-\I QCx^3}$. This leads to $A_0C\rightarrow\e^{-\I QCx^3}E\e^{\I QCx^3}$, 
which is $x^3$ dependent.
Choosing $x^0=0$ as the lower integration bound in the expression for $A_3$, we find for the Wilson loops the 2 different results,
\be
W&=&\tr~\e^{\I[Q,\sint_0^{y^0}dtE(t)]y^3}~~~~~~~\mathrm{and}\nn
W&=&\tr~\e^{-\I QCy^3}\e^{\I\int^{y^0}_0 dtE(t)/C}\e^{\I QCy^3}
\e^{\I\int_0^{y^0}dt E(t)/C} .
\nonumber
\ee


In mixed representation, in a purely time-dependent background, the equation 
of motion for the {\it Klein--Gordon} propagator 
$\tilde S=\tilde S(x_0,y_0,\vec p)$ reads
\be
(\partial_0^2-\I\dot A_0-2\I A_0\partial^0+2A_jp^j-A{\cdot}A+\omega^2)
\tilde S
=
\delta^{(1)},
\label{kgmixed}
\ee
where 
$\delta^{(1)}=\delta(x_0-y_0)$, 
$j\in\{1;2;3\}$, and $\omega^2=|\vec p|^2+m^2$.
%
%
For configuration (\ref{config3}) this becomes,
\be
[\partial_0^2+(p_3-E_3x^0)^2+m_\perp^2]
\tilde S
=
\delta^{(1)} ,
\ee
where $m_\perp^2=|\vec p_\perp|^2+m^2$.
A decomposition into eigenvectors of $E_3$ leads to 
\be
[\partial_0^2+(p_3-e_nx^0)^2+m_\perp^2]
\tilde S_n
=
P_n
\delta^{(1)} ,
\ee
where $E_3|n\rangle=e_n|n\rangle$, $\langle n|m\rangle=\delta_{nm}$,
$P_n=|n\rangle\langle n|$ and $\tilde S_n=P_n\tilde S$.
The homogeneous solutions to this differential equation are,
\be
M_l(x^0)
=
t^{-\sfrac{1}{2}}
M_{-\sfrac{\I m_\perp^2}{4e_n},-\sfrac{(-1)^l}{4}}(\I e_nt^2) ,
\label{whittakerkg}
\ee
where $l\in\{1;2\}$, $t=x^0-p_3/e_n$, and $M_{\kappa,\mu}(z)$ is a Whittaker
function. \{See Eqs.~(13.1.31) and (13.1.32) in \cite{AS}.\} With the
boundary conditions $\tilde S=0$ and $\dot{\tilde S}=1$ at $x_0=y_0$, we find 
for the retarded solution,
\be
\tilde S
=
\sum_nP_n
\frac
{
M_1(x^0)M_2(y^0)
-
(1\leftrightarrow 2)
}{
\dot M_1(y^0)M_2(y^0)-(1\leftrightarrow 2)
}
\theta^{(1)} ,
\label{dprop3}
\ee
where $\theta^{(1)}=\theta(x_0-y_0)$ stands for the Heaviside step function. 
The denominator contains a known Wronskian and evaluates to i.
\{See Eqs.~(13.1.34), (13.1.32), (13.1.33) and (13.1.22) in \cite{AS}.\}
In the limit of large $t$, Eq.~(\ref{whittakerkg}) becomes \{see Eqs.~(13.1.32) and (13.5.1) in \cite{AS}\}
\be
M_l
&\rightarrow&
\I^{\sfrac{c_l}{2}}\Gamma(c_l)\e^{-\sfrac{\pi m_\perp^2}{8e_n}}
t^{-\sfrac{1}{2}}
\times\nn&&\times
\left[
\frac
{
(e_nt^2)^{-\sfrac{\I m_\perp^2}{4e_n}}\e^{-\sfrac{\I}{2}e_nt^2}
\I^{\sfrac{c_l}{2}}
}{
\Gamma(\sfrac{c_l}{2}-\sfrac{\I m_\perp^2}{4e_n})
}
+
\mathrm{c.c.}
\right] ,
\ee
where $c_1=3/2$ and $c_2=1/2$. In the previous expression, we can already see the typical exponential $m_\perp$ behaviour of the pair 
production rate. 


For configuration (\ref{config03}), Eq.~(\ref{kgmixed}) becomes
\be
[\partial_0^2-2\I a_0\partial^0-(a_0)^2+(p_3-a_3)^2+m_\perp^2]
\tilde{\underline{S}}
=
\delta^{(1)} .
\label{kg03}
\ee
Let us continue with $SU(2)$ [at least an $SU(2)$ subgroup], the 
generalisation to higher gauge groups being straightforward. Define
\be
[\partial_0^2+2\I a_0\partial^0-(a_0)^2+(p_3+a_3)^2+m_\perp^2]
\tilde{\underline{s}}
=
\tilde{\underline{S}} .
\nonumber
\ee
Then, from Eq.~(\ref{kg03}), assuming $\{a_0,a_3\}=0$,
\be
\{[\partial_0^2-(a_0)^2+(a_3)^2+\omega^2]^2
+&&\nn+
4(a_0)^2\partial_0^2
-
4(p_3)^2(a_3)^2\}
\tilde{\underline{s}}
&=&
\delta^{(1)} .
\ee
The exponential ansatz $\tilde{\underline{s}}\sim\e^{\lambda x^0}$ yields
the 4 values, 
\be
\lambda^2_\pm
&=&
-[\omega^2+(a_0)^2+(a_3)^2]
\pm\nn&&\pm
2\sqrt{(a_0)^2(a_3)^2+(a_0)^2\omega^2+(p_3)^2(a_3)^2} .
\nonumber
\ee
For $[\omega^2-(a_0)^2+(a_3)^2]^2<4(p_3)^2(a_3)^2$, this corresponds to 2
oscillatory, 1 exponentially decaying, and 1 exponentially growing mode;
otherwise, the behaviour is purely oscillatory.
For comparison, repeating the same steps for a magnetic field
$F_{12}^a=f^{abc}a_1^ba_2^c$ yields
\be
\lambda^2_\pm
=
-[(a_1)^2+(a_2)^2+\omega^2]
\pm
2
\sqrt{
(a_1)^2(p_1)^2+(a_2)^2(p_2)^2
} ,
\nonumber
\ee
implying always purely oscillatory solutions.


In mixed representation the {\it Dirac} equation is given by
\be
(\I\gamma^0\partial_0-\gamma^jp_j+\SSH{A}-m)\tilde G=\delta^{(1)},
\ee
where $\tilde G=\tilde G(x_0,y_0,\vec p)$. With the help of
\be
-(\I\gamma^0\partial_0-\gamma^jp_j+\SSH{A}+m)\tilde g=\tilde G.
\label{deqsq}
\ee
we obtain the squared Dirac equation,
\be
[\partial_0^2-2\I A^0\partial_0+2p_jA^j-\I\gamma^0\SSH{\dot
A}-\SSH{A}\SSH{A}+\omega^2]
\tilde g
=
\delta^{(1)} .
\label{deqmixed}
\ee
%
%
For configuration (\ref{config3}) this becomes
\be
[\partial_0^2+(p_3-E_3x^0)^2+m_\perp^2-\I\gamma^0\gamma^3E_3]
\tilde g
=
\delta^{(1)} .
\label{deq3}
\ee
We continue by carrying out a decomposition with the
projectors $P_\pm=(1\pm\gamma^0\gamma^3)/2$ and $P_n$,
\be
[\partial_0^2+(p_3-e_nx^0)^2+m_\perp^2\mp\I e_n]
\tilde g_\pm^n
=
P_nP_\pm\delta^{(1)},
\ee
where $\tilde g_\pm^n=P_nP_\pm\tilde g$. Up to the substitutions
$m_\perp^2\rightarrow m_\perp^2\mp\I e_n$, $\tilde g_\pm^n$ are the same as 
Eq.~(\ref{dprop3}), 
\be
\I\tilde g
=
\sum_{n,\pm}P_\pm P_n
[
M_1^\pm(x_0)M_2^\pm(y_0)
-
(1\leftrightarrow 2)
]
\theta^{(1)}.
\label{deqsq3}
\ee
In the limit of large $t$, $M_l^\pm$ become \{see Eqs.~(13.1.32)
and (13.5.1) in \cite{AS}\}
\be
M_l^\pm
&\rightarrow&
\I^{\sfrac{c_l}{2}\mp\sfrac{1}{4}}\Gamma(c_l)\e^{-\sfrac{\pi m_\perp^2}{8e_n}}
(e_nt^2)^{-\sfrac{1}{4}}
\times\\&\times&
\left[
\frac
{
(e_nt^2)^{\pm\sfrac{1}{4}-\sfrac{\I m_\perp^2}{4e_n}}\e^{-\sfrac{\I}{2}e_nt^2}
\I^{\sfrac{c_l}{2}}
}{
\Gamma(\sfrac{c_l}{2}\pm\sfrac{1}{4}-\sfrac{\I m_\perp^2}{4e_n})
}
+
\binom{\mathrm{c.c.}~\&}{\pm\leftrightarrow\mp}
\right] .
\nonumber
\ee
At the end, the Dirac propagator is obtained by putting Eq.~(\ref{deqsq3}) into
Eq.~(\ref{deqsq}). At late times, the Dirac operator in Eq.~(\ref{deqsq}) is
dominated by the gauge field term, which grows linearly and the derivative
term, which, when acting on the Gaussian in time in the previous equation
also generates an extra factor of time. Hence, the dominant components of
$\tilde G$ are growing approximately like the square root of time. 
If we use $\tilde G$ to construct the fermion current
$\bar{\psi}\gamma^\mu\psi$ this factor appears twice and the current grows
linearly in time. Therefore, one talks of a constant pair production rate in
this field configuration.


For configuration (\ref{config03}), Eq.~(\ref{deqmixed}) becomes
\be
&&[\partial_0^2-2\I a^0\partial_0-(a_0)^2
+\nn&&+
(p_3-a_3)^2+m_\perp^2-\I\gamma^0\gamma^3E_3]
\tilde{\underline{g}}
=
\delta^{(1)} .
\label{deq03}
\ee
We carry out the same decomposition with the projectors $P_\pm$ as before for
Eq.~(\ref{deq3}), 
\be
&&[\partial_0^2-2\I a^0\partial_0-(a_0)^2
+\nn&&+
(p_3-a_3)^2+m_\perp^2\mp\I E_3]
\tilde{\underline{g}}_\pm
=
P_\pm\delta^{(1)},
\ee
where $\tilde{\underline{g}}_\pm=P_\pm\tilde{\underline{g}}$. Define
\be
&&[\partial_0^2+2\I a^0\partial_0-(a_0)^2
+\nn&&+
(p_3+a_3)^2+m_\perp^2\pm\I E_3]
\tilde{\underline{\Gamma}}_\pm
=
\tilde{\underline{g}}_\pm.
\label{deqfourth}
\ee
Then, with pairwise anticommuting $a_0$, $a_3$ and $E_3$,
\be
&&\{[\partial_0^2-(a_0)^2+(a_3)^2+\omega^2]^2
+\\&&+
4(a_0)^2\partial_0^2
-
4(p_3)^2(a_3)^2+(E_3)^2\}
\tilde{\underline{\Gamma}}_\pm
=
P_\pm\delta^{(1)} .
\nonumber
\ee
From the exponential ansatz $\tilde\Gamma_\pm\sim\e^{\mu x^0}$ we get
\be
\mu_\pm^2
&=&
-[\omega^2+(a_0)^2+(a_3)^2]
\pm\nn&&\pm
\sqrt{4[(a_0)^2(a_3)^2+(a_0)^2\omega^2+(p_3)^2(a_3)^2]-(E_3)^2},
\nonumber
\ee
which leads to a purely oscillatory behaviour, as does the analogous result
for a magnetic $B_3$ field,
\be
\mu^2_\pm
&=&
-[(a_1)^2+(a_2)^2+\omega^2]
\pm\nn&&\pm
\sqrt{
4[(a_1)^2(p_1)^2+(a_2)^2(p_2)^2]+(B_3)^2
} .
\ee


{\it In conclusion}, there exist non-Abelian field tensors that can be realised
by different gauge field configurations that are not linked by gauge
transformations, i.e., that are not gauge equivalent. Under these 
circumstances the covariant derivative carries more information than its 
commutator, the field tensor. In most of the gauge-inequivalent
configurations leading to the same field tensor there exist observables
(gauge invariant quantities) that cannot be expressed exclusively in terms
of the field tensor. Here, we have demonstrated this explicitly for various 
field tensors that allow for gauge-inequivalent gauge field realisations.
As examples we have picked static field tensors, electric or magnetic, and
purely time-dependent configurations. Concretely, we showed that a direct 
gauge transformation of different gauge field configurations into each other
cannot be found despite the fact that they yield the same field tensor;
further, that for these different configurations the corresponding Wilson 
loops and Yang--Mills currents differ, as do the Klein--Gordon and Dirac 
propagators. For example, while the induced fermion current in the 
Abelian-like realisation for a static electric field exhibits asymptotically 
linear growth with time, which leads to the rate interpretation of the result, the propagators in the
genuinely non-Abelian realisation possess only purely oscillatory modes. In
the latter realisation, the scalar propagator can also feature exponentially 
growing and decaying modes in the presence of an electric field.
A particular quantity that cannot be expressed in terms of a Wu--Yang
ambiguous field tensor is the Yang--Mills current. It is exactly the 
covariantly constant case, where this current 
vanishes, which explains why the effective actions for scalars or fermions in 
this configuration can be expressed in terms of the field tensor alone.
For $\det\mathbbm{F}\neq0$ there is no Wu--Yang ambiguity and $A^a_\mu$ can 
be expressed in terms of $F^a_{\mu\nu}$ and therefore, all invariants and
observables.
 
In the worldline approach \cite{WL} to effective actions all these differences
discussed above reflect in the precession of the colour as described by Wong's equation
\cite{Wong:1970fu}.

The above facts may have consequences for the flux-tube picture \cite{Casher:1978wy} for ultrarelativistic collisions, which features static chromoelectric fields. Depending on how the latter is realised, by configuration (\ref{config3}) or (\ref{config03}) [or if, e.g., a decaying field is assumed by configuration (\ref{config3t}) or (\ref{config03t})], the particle yields differ. 
$J_0$ can serve to distinguish between the realisations.

As mentioned in \cite{Dietrich:2009an}, also Coulomb fields have $\det\mathbbm{F}=0$. Boosted onto the light-cone, i.e., as Weizs\"acker--Williams fields, they are used in the colour glass condensate framework \cite{McLerran:1993ni} to model the initial conditions of ultrarelativistic collisions.


{\it Acknowledgments}.
The author would like to acknowledge discussions with 
J.~Reinhardt, 
F.~Sannino, 
K.~Socha,
and
K.~Tuominen.
His work was supported by the Danish Natural Science Research Council. 



\begin{thebibliography}{99}

\bibitem{DDD:ConstantFields}
Already choosing an ensemble of pure gauge configurations
leads to unexpected phenomena. See, e.g.,
  P.~Hoyer and S.~Peigne,
  arXiv:hep-ph/0304010;
  JHEP {\bf 0412} (2004) 051
  [arXiv:hep-ph/0410235];
  D.~D.~Dietrich and S.~Hofmann,
  Phys.\ Lett.\  B {\bf 632} (2006) 439
  [arXiv:hep-ph/0506210];
  D.~D.~Dietrich,
  Phys.\ Rev.\  D {\bf 74} (2006) 065023
  [arXiv:hep-ph/0507112];
  D.~D.~Dietrich, P.~Hoyer, M.~Jarvinen and S.~Peigne,
  JHEP {\bf 0703} (2007) 105
  [arXiv:hep-ph/0608075].

\bibitem{ConstantFieldHistory}
F.~Sauter,
Z.\ Phys.\ {\bf 69} (1931) 742;
W.~Heisenberg and H.~Euler,
Z.\ Phys.\ {\bf 98} (1936) 714;
V. Weisskopf,
Kgl.\ Danske Videnskab.\ Selskabs.\ Mat.-fys.\ Medd.\ {\bf 14} No.~6 (1936);
J.~Schwinger,
Phys.\ Rev.\ {\bf 82} (1951) 664.

\bibitem{ELI}
  H.~Gies,
  arXiv:0812.0668 [hep-ph];
  G.~V.~Dunne,
  arXiv:0812.3163 [hep-th].

\bibitem{Dunne:2004nc}
  G.~V.~Dunne,
  arXiv:hep-th/0406216.

\bibitem{nabeffact}
  M.~R.~Brown and M.~J.~Duff,
  Phys.\ Rev.\  D {\bf 11} (1975) 2124;
  M.~J.~Duff and M.~Ramon-Medrano,
  Phys.\ Rev.\  D {\bf 12} (1975) 3357;
  I.~A.~Batalin, S.~G.~Matinyan and G.~K.~Savvidy,
  Sov.\ J.\ Nucl.\ Phys.\  {\bf 26} (1977) 214
  [Yad.\ Fiz.\  {\bf 26} (1977) 407];
  G.~K.~Savvidy,
  Phys.\ Lett.\  B {\bf 71} (1977) 133;
  S.~G.~Matinyan and G.~K.~Savvidy,
  Nucl.\ Phys.\  B {\bf 134} (1978) 539;
  N.~K.~Nielsen and P.~Olesen,
  Nucl.\ Phys.\  B {\bf 144} (1978) 376;
  Phys.\ Lett.\  B {\bf 79} (1978) 304;
  G.~C.~Nayak and P.~van Nieuwenhuizen,
  Phys.\ Rev.\  D {\bf 71} (2005) 125001
  [arXiv:hep-ph/0504070];
  G.~C.~Nayak,
  Phys.\ Rev.\  D {\bf 72} (2005) 125010
  [arXiv:hep-ph/0510052];
  F.~Cooper and G.~C.~Nayak,
  Phys.\ Rev.\  D {\bf 73} (2006) 065005
  [arXiv:hep-ph/0511053];
  F.~Cooper, J.~F.~Dawson and B.~Mihaila,
  Phys.\ Rev.\  D {\bf 78} (2008) 117901
  [arXiv:0811.3905 [hep-ph]].

\bibitem{Invariants}
  R.~Roskies,
  Phys.\ Rev.\  D {\bf 15} (1977) 1722;
  J.~Anandan and R.~Roskies,
  Phys.\ Rev.\  D {\bf 18} (1978) 1152;
  J.\ Math.\ Phys.\  {\bf 19} (1978) 2614;
  G.~C.~Nayak and R.~Shrock,
  Phys.\ Rev.\  D {\bf 77} (2008) 045008
  [arXiv:0711.2759 [hep-th]].

\bibitem{WY}
  T.~T.~Wu and C.~N.~Yang,
  Phys.\ Rev.\  D {\bf 12} (1975) 3843.

\bibitem{WY:Det4d}
  S.~Deser and F.~Wilczek,
  Phys.\ Lett.\  B {\bf 65} (1976) 391;
  R.~Roskies,
  Phys.\ Rev.\  D {\bf 15} (1977) 1731;
  M.~Calvo,
  Phys.\ Rev.\  D {\bf 15} (1977) 1733;
  S.~R.~Coleman,
  Phys.\ Lett.\  B {\bf 70} (1977) 59.

\bibitem{B}
  S.~Deser and C.~Teitelboim,
  Phys.\ Rev.\  D {\bf 13} (1976) 1592.
  M.~B.~Halpern,
  Nucl.\ Phys.\  B {\bf 139} (1978) 477;
  D.~Z.~Freedman and R.~R.~Khuri,
  Phys.\ Lett.\  B {\bf 329} (1994) 263
  [arXiv:hep-th/9403031];
  D.~D.~Dietrich,
  arXiv:0704.1828 [hep-th];
  arXiv:0804.0904 [hep-th].

\bibitem{Brown:1979bv}
  L.~S.~Brown and W.~I.~Weisberger,
  Nucl.\ Phys.\  B {\bf 157} (1979) 285
  [Erratum-ibid.\  B {\bf 172} (1980) 544].

\bibitem{DDD:Propagators}
See also
  D.~D.~Dietrich,
  Phys.\ Rev.\  D {\bf 68} (2003) 105005
  [arXiv:hep-th/0302229];
  Phys.\ Rev.\  D {\bf 70} (2004) 105009
  [arXiv:hep-th/0402026];
  Phys.\ Rev.\  D {\bf 71} (2005) 045005
  [arXiv:hep-ph/0411245];
  Phys.\ Rev.\  D {\bf 74} (2006) 085003
  [arXiv:hep-th/0512026];
  G.~C.~Nayak, D.~D.~Dietrich and W.~Greiner,
  arXiv:hep-ph/0104030.
  D.~D.~Dietrich, G.~C.~Nayak and W.~Greiner,
  arXiv:hep-ph/0009178;
  Phys.\ Rev.\  D {\bf 64} (2001) 074006
  [arXiv:hep-th/0007139].

\bibitem{AS}
  M.~Abramowitz and I.~A.~Stegun,
  ``Handbook of Mathematical Functions",
  Dover, 1964, New York.

\bibitem{WL}
  M.~J.~Strassler,
  Nucl.\ Phys.\  B {\bf 385} (1992) 145
  [arXiv:hep-ph/9205205];
  C.~Schubert,
  Phys.\ Rept.\  {\bf 355} (2001) 73
  [arXiv:hep-th/0101036];
  H.~Gies and K.~Klingmuller,
  Phys.\ Rev.\  D {\bf 72} (2005) 065001
  [arXiv:hep-ph/0505099];
  G.~V.~Dunne and C.~Schubert,
  Phys.\ Rev.\  D {\bf 72} (2005) 105004
  [arXiv:hep-th/0507174];
  G.~V.~Dunne, Q.~h.~Wang, H.~Gies and C.~Schubert,
  Phys.\ Rev.\  D {\bf 73} (2006) 065028
  [arXiv:hep-th/0602176];
  D.~D.~Dietrich and G.~V.~Dunne,
  J.\ Phys.\ A  {\bf 40} (2007) F825
 [arXiv:0706.4006 [hep-th]];
  G.~V.~Dunne,
  J.\ Phys.\ A  {\bf 41} (2008) 164041.

\bibitem{Wong:1970fu}
  S.~K.~Wong,
  Nuovo Cim.\  A {\bf 65S10} (1970) 689.
  
\bibitem{Casher:1978wy}
  A.~Casher, H.~Neuberger and S.~Nussinov,
  Phys.\ Rev.\  D {\bf 20} (1979) 179.
  
\bibitem{Dietrich:2009an}
  D.~D.~Dietrich,
  Phys.\ Rev.\  D {\bf 79} (2009) 107703
  [arXiv:0904.0820 [hep-th]].
  
\bibitem{McLerran:1993ni}
  L.~D.~McLerran and R.~Venugopalan,
  Phys.\ Rev.\  D {\bf 49} (1994) 2233
  [arXiv:hep-ph/9309289];
  Phys.\ Rev.\  D {\bf 49} (1994) 3352
  [arXiv:hep-ph/9311205];
  Phys.\ Rev.\  D {\bf 50} (1994) 2225
  [arXiv:hep-ph/9402335].

\end{thebibliography}
\end{document}